\documentclass[manuscript]{aastex}

\shorttitle{Close Pairs in the Millennium Galaxy Catalogue}
\shortauthors{De~Propris et al.}

\begin{document}


\title{The Millennium Galaxy Catalogue: Dynamically close
       pairs of galaxies and the global merger rate}


\author{Roberto~De~Propris}
\affil{H.~H.~Wills Physics Laboratory, University of Bristol,\\
       Tyndall Avenue, Bristol, BS8 1TL, United Kingdom}
\email{R.DePropris@bristol.ac.uk}

\author{Jochen~Liske}
\affil{European Southern Observatory, Karl-Schwarzschild-Stra{\ss}e 2,\\
       85748 Garching b.~M{\"u}nchen, Germany}
\email{jliske@eso.org}

\author{Simon~P.~Driver\altaffilmark{1} and Paul D. Allen}
\affil{Research School of Astronomy and Astrophysics, Australian
       National University, \\
       Cotter Road, Weston, ACT, 2611, Australia}
\email{spd,paul@mso.anu.edu.au}

\and

\author{Nicholas~J.~G.~Cross}
\affil{Institute for Astronomy, The University of Edinburgh,
       Royal Observatory, Blackford Hill, Edinburgh, EH9 3HJ,
       United Kingdom }
\email{njc@roe.ac.uk}


\altaffiltext{1}{PPARC Visiting Fellow, University of Bristol}


\begin{abstract}

We derive the number of dynamically close companions per galaxy
($N_c$) and their total luminosity ($L_c$) for galaxies in the
Millennium Galaxy Catalogue: $N_c$ is similar to the fraction of
galaxies in close pairs and is directly related to the galaxy
merger rate. We find $N_c=0.0174 \pm 0.0015$ and $L_c=(252 \pm 30) 
\times 10^6$ $L_{\odot}$ for galaxies with $-22 < M_B -5 \log h < -19$ 
with $<z>=0.123$ and $N_c=0.0357 \pm 0.0027$, $L_c= (294 \pm 31) 
\times 10^6 \; L_{\odot}$ for galaxies with $-21 < M_B -5 \log h < -18$,
with $<z>=0.116$. The integrated merger rate to $z=1$ for both samples is 
about 20 \%, but this depends sensitively on the fraction of kinematic
pairs that are truly undergoing a merger (assumed here to be 50\%),
the evolution of the merger rate (here as $(1+z)^3$) and the adopted
timescale for mergers (0.2 and 0.5 Gyr for each sample, respectively). 
Galaxies involved in mergers tend to be marginally bluer than 
non-interacting galaxies and show an excess of both early-type and 
very late-type objects and a deficiency of intermediate-type spirals. 
This suggests that interactions and mergers partly drive the star formation 
and morphological evolution of galaxies.

\end{abstract}




\section{Introduction}

In the Cold Dark Matter (CDM) framework, galaxies are assembled
gradually via a process of hierararchical mergers, where increasingly
more massive subunits coalesce to produce today's luminous giant
galaxies (e.g., \citealt{cole00} and references therein): the history
of mass assembly of galaxies is then reflected in the global merger 
rate and in its evolution with lookback time. Mergers and close interactions 
may also play a role in the morphological transformation of galaxies (see, 
for instance, \citealt{hernandez05,scannapieco03,steinmetz02,
junqueira98,mihos96}), triggering of starbursts \citep{nikolic04,alonso04,
barton03,bergvall03,lambas03,tissera02,barton00,donzelli97} and fuelling 
of active galactic nuclei (e.g., \citealt{sanchez03,canalizo01} ). 

Traditionally, the observational route to measuring the merger
rate has been the conventional pair fraction, under the assumption
that sufficiently close galaxies will result in a merger over
relatively short timescales \citep{zepf89,burkey94,carlberg94,
woods95,yee95,patton97,wu98,bundy04};  however, limited redshift 
information makes the derived pair fraction dependent on the mean 
galaxy density and the correlation function, and all such estimates 
are affected by contamination from unphysical pairs. 

In order to address this issue, it is sometimes required that pairs also 
show evidence of interactions \citep{neuschafer97,lefevre00,hashimoto00,xu04}: 
on the other hand, this introduces an element of subjectivity in the analysis, 
as a threshold of morphological disturbance must be chosen for objects to be 
considered parts of pairs. Furthermore, a number of true pairs may not show 
evidence of tidal tails or other deviations from symmetry (because of low 
surface brightness, cosmological dimming and morphological $k$-corrections, 
for example; see discussion in \citealt{mihos95}).

The fraction of galaxies in kinematic pairs (i.e. both spatially and 
dynamically close) yields a more rigorous estimate of the local pair 
fraction and global merger rate, although the method requires highly 
complete redshift information and accurate control of systematic biases 
arising from the flux-limited nature of redshift surveys and from 
incompleteness. At least for the local universe, where large redshift 
samples are currently available, this approach is now feasible
and was first used by Patton et al. (2000, hereafter P00) who measured the 
number of dynamically close companions per galaxy within the range $-21 
< M_B < -18$ (a statistic akin to the pair fraction)  using data from the 
Second Southern Sky Redshift Survey \citep{dacosta98}, and established the 
necessary mathematical formalism (see next section for a brief summary).

The two large redshift surveys now available, the 2dF Galaxy Redshift
Survey (2dFGRS; \citealt{colless01}), and the Sloan Digital Sky
Survey (SDSS; \citealt{york00}), should provide a more extensive
sample for such studies. Of these, the 2dFGRS is currently the
largest and most complete publicly available dataset. However,
because of its complicated selection function at small angular
separation, which is related to restrictions on fiber placement and the
survey tiling strategy, this survey is not well suited to a study of close 
pairs \citep{hawkins03}. In this paper we choose to study a smaller, but 
more complete dataset, which is more appropriate for the purpose of measuring 
the number of close companions per galaxy and the local merger rate.

The Millennium Galaxy Catalogue, hereafter MGC \citep{liske03},
covers a $35'$ wide and $72.2^{\circ}$ long equatorial strip, coinciding
with the northern strip of the 2dFGRS and with the SDSS Data Release
1 region \citep{abazajian03}, which provide $\sim 50\%$ of the
MGC redshifts as well as $ugriz$ photometry. The MGC has been reimaged 
in the $B$ band to deep surface brightness limits (26 mag arcsec$^{-2}$) 
in order to derive accurate structural parameters, and additional redshifts
have been collected to bring the redshift completeness to 99.79\% for
galaxies with $B < 19.2$ and 96.05\% for galaxies with $B < 20$.
The total sample includes 10,095 galaxies to $B=20$ \citep{driver05}. 
Because of its high completeness and photometric accuracy \citep{cross04}, 
the MGC is well suited to a study of the local merger rate via close pairs 
and to an analysis of the properties of the dynamically paired galaxies.
Even the MGC, however, suffers from some incompleteness at small
separations, for which we will need to make allowance in our analysis.

We describe the procedures used to derive the number of close
companions per galaxy in the next section, apply these techniques to
the MGC in section 3 and present the results and our discussion in
section 4, dealing with theoretical interpretation of the data and
with the properties of galaxies in pairs. We adopt a cosmology with
$\Omega_M=0.3$, $\Omega_{\Lambda}=0.7$ and H$_0=100$ km s$^{-1}$
Mpc$^{-1}$ and all absolute magnitudes used should be understood with
reference to this cosmology.

\section{Methodology}

We follow the method of P00 and the reader is referred to that paper 
(and Patton et al. 2002, hereafter P02) for a more complete discussion. 
Here we present a summary of the technique and repeat some essential 
points.

Consider a primary sample of $N_1$ galaxies and a secondary sample of
$N_2$ galaxies. We are interested in determining the number of
galaxies from the secondary sample that are dynamically 'close' to a
primary galaxy, per primary galaxy. This statistic, $N_c$, is similar
to the fraction of galaxies in close pairs and is in fact identical to
it for a volume-limited sample which contains only close pairs but no
triplets or higher order $N$-tuples. We define galaxies to be
dynamically close if they have a projected separation $5 < r_p <
20$~kpc (the inner limit is set to avoid confusion with star formation
knots or HII regions) and if they have a relative velocity of less
than 500~km~s$^{-1}$. Simulations show that such pairs would result in
a merger on timescales of 10$^9$ years \citep{toomre77,barnes88} and
therefore $N_c$ is directly related to the galaxy merger rate.
These requirements have been conventionally used in the literature (e.g., 
\citealt{carlberg94,yee95,carlberg00}) on close pair statistics.

P00 have shown that for a flux-limited sample (like
the MGC) the number of close companions per galaxy is best computed as:
\begin{equation}
\label{nc}
N_c={{\sum_i^{N_1} w_{N_1}^i N_{c_i}} \over {\sum_i^{N_1} w_{N_1}^i}}
\end{equation}
and the total companion luminosity as:
\begin{equation}
\label{lc}
L_c={{\sum_i^{N_1} w_{L_1}^i L_{c_i}} \over {\sum_i^{N_1} w_{L_1}^i}}.
\end{equation}
$N_{c_i}$ and $L_{c_i}$ are the number and luminosity of galaxies from the
secondary sample that are dynamically close to the $i$th primary
galaxy. They are given by:
\begin{equation}
\label{nci}
N_{c_i} = \sum_j w_{N_2}^j = \sum_j \frac{w_{b_2}^j w_{v_2}^j}{S_N(z_j)}
\end{equation}
and
\begin{equation}
\label{lci}
L_{c_i} = \sum_j w_{L_2}^j L_j = \sum_j \frac{w_{b_2}^j
  w_{v_2}^j}{S_L(z_j)} L_j,
\end{equation}
where the sums run over those secondary galaxies that fulfill the
criteria of being dynamically close to the $i$th primary galaxy.

The weights that are being applied to the secondary sample, $w_{N_2}$
and $w_{L_2}$, have three components. The first factor corrects for
the change of density of the secondary galaxies as a function of
redshift due to the apparent flux limit of the sample: for a
low-redshift primary galaxy we will find more close companions on
average than for a higher redshift one because the companions can be
drawn from a wider range of luminosities (cf.\ Figure 1). The
appropriate (inverse) correction factors are given by the selection
functions $S_N(z)$ and $S_L(z)$ which are defined in terms of ratios
of densities in flux limited vs.\ volume limited samples:
\begin{equation}
S_N (z) = {{\int^{M_{lim}(z)}_{M_{bright}} \Phi (M) dM} \over
            {\int^{M_2}_{M_{bright}} \Phi (M) dM}}
\end{equation}
and
\begin{equation}
S_L (z) = {{\int^{M_{lim}(z)}_{M_{bright}} \Phi (M) L(M) dM} \over
            {\int^{M_2}_{M_{bright}} \Phi (M) L(M) dM}}.
\end{equation}
$\Phi(M)$ denotes the luminosity function for which we adopt the
latest MGC values from \cite{driver05} ($\Phi^*=0.0177 \; {\rm
Mpc}^{-3}$, $M^*=-19.60$ and $\alpha=-1.13$; $B$ band). $M_{lim}$ 
is the redshift dependent absolute magnitude limit:
\begin{equation}
M_{lim}(z) = {\rm max}(M_{faint},m-5 \log d_L(z)-25-k(z)-e(z)),
\end{equation}
where $m$ is the apparent magnitude limit of the survey ($B=20$), $d_L$ 
is the luminosity distance, $k(z)$ is the maximum $k$-correction
at redshift $z$ (P02) and $e(z)$ is of the form $-0.75 \times 2.5 \log (1+z)$, 
which corresponds to a passive luminosity evolution scenario with no mergers:
this form yields the same results (to within 0.01) of the $e(z)=-0.7z$ 
assumed by P02. Here, $M_{bright}$ and $M_{faint}$ are magnitude limits 
introduced to take account of the fact that the clustering properties of 
galaxies vary with luminosity \citep{norberg02} and therefore we must 
only use galaxies in a limited luminosity range so that the clustering 
length does not vary too strongly.

In essence, the factors $S_N$ and $S_L$ in equations \ref{nci} and
\ref{lci} correct the secondary sample from a flux limited sample in
the range $M_{bright} < M < M_{lim}(z)$ to a hypothetical volume
limited sample in the range $M_{bright} < M < M_2$.

The other two components of the $w_{N_2}$ weights correct for boundary
effects. $w_{b_2}$ corrects for the fact that some fraction, $1 -
f_b^i$, of the $r_p$ area around the $i$th primary galaxy may lie
outside the effective survey area: $w_{b_2}^j = 1 / f_b^i$. For
primary galaxies lying close to the survey boundaries in redshift
space (i.e.\ within 500 km~s$^{-1}$) $w_{v_2}$ corrects for possible
companions beyond those limits: we ignore all companions between the
primary galaxy and the redshift boundary and adopt $w_{v_2}^j=2$ for
those companions lying away from the boundary.

Finally we note that in equations \ref{nc} and \ref{lc} we also
apply weights to the primary sample. The purpose of the $w_{N_1}$ and
$w_{L_1}$ weights is to minimise the errors on $N_c$ and $L_c$. P00
have shown that they should be chosen as:
\begin{equation}
w_{N_1}^i= w_{b_1}^i w_{v_1}^i S_N(z_i)
\end{equation}
\begin{equation}
w_{L_1}^i= w_{b_1}^i w_{v_1}^i S_L(z_i),
\end{equation}
where $w_{b_1}^i = f_b^i$ and $w_{v_1}^i = 0.5$ (i.e. the reciprocals
of the weights for the secondary sample) for primary galaxies close to 
the redshift boundaries.

\section{Application to the MGC}

We now apply these techniques to the MGC. Figure 1 shows the absolute
magnitude vs.\ redshift for all MGC galaxies with $B < 20$: this
includes 9696 galaxies with redshifts out of 10,095, with 96.05\%
redshift completeness.

The mean absolute magnitude of the MGC, weighted according to equations
(8) and (9) above, is $M_B=-18.5$. We then choose to analyze two subsets
of the MGC data, one with $M_{bright}=-22$ and $M_2=-19$ and another with
$M_{bright}=-21$ and $M_2=-18$ (to which ranges the results are normalized). 
Selection lines for both samples are drawn in Figure 1. For these samples, 
we use $M_{faint}=-18$ and $-17$ (respectively); the brighter sample contains 
5756 galaxies, while the fainter one includes 6492 galaxies. The magnitude limits 
chosen here (as in other studies) imply that we are mostly concerned 
with major mergers between galaxies of approximately similar luminosity.
As in P02 some galaxies are brighter than the apparent magnitude limit but blue
and therefore lie `below' the selection line in Figure 1; these galaxies
are excluded from the analysis to ensure that galaxies of all spectral
types have an equal probability of being included within the samples.

For the bright sample ($-22 < M_B < -19$) we find a total of 137 dynamically
close companions in 69 pairs\footnote{One of the pairs contributes only a
single companion because $r_p$ is just below 20 kpc at the redshift of one 
of the galaxies but just above 20 kpc at the redshift of the other}, where 
two galaxies appear in two pairs. Of these companions, 60 lie in the 
magnitude range $-19 < M_B < -18$ (i.e. between $M_2$ and $M_{faint}$
for this sample. For the faint sample ($-21 < M_B < -18$) we find 176 
companions in 89 pairs, with one triple and where two galaxies appear in 
two pairs each. Of these, 27 lie between $M_2$ and $M_{faint}$. This is
similar to P00 in that galaxies between these two limits do not contribute
significantly to the pair fraction for the $\sim L^*$ galaxies; however,
these objects are much more significant contributors for the brighter
sample. This may be evidence that the brighter galaxies tend to have fainter
companions.

Applying equations 1 and 2 we find $N_c=0.0147 \pm 0.0013$ and 
$L_c={213 \pm 25} \times 10^6 \; L_{\odot}$ for galaxies with 
$-22 < M_B < -19$, with  mean $z = 0.124$ and $N_c=0.0301 \pm 0.0023$, 
with $L_c=(248 \pm 26) \times 10^6 \; L_{\odot}$ for galaxies with 
$-21 < M_B < -18$ with mean $z=0.116$, where the errors are derived 
by jackknife resampling of the data \citep{efron82}.

Figure 2 shows postage stamps for a random selection of pairs; about half or
more of these galaxies show clear signs of interaction such as
disturbed morphologies, rings and tidal features, demonstrating that
our sample indeed includes {\it bona fide} physical pairs. The full
list of companions provides a useful objectively selected sample of
galaxies in the early phase of a merger. Table 1 lists all unique
pairs found in MGC; in order the columns show: the MGC ID of the
primary galaxy, the IDs of the other pair members, the RA and Dec.
(2000) for the primary galaxy and its radial velocity.

However, we know that the MGC has some redshift incompleteness. We
proceed to estimate the number of missed pairs in the following
way. We repeat our analysis including galaxies without redshifts and
using only the $r_p$ criterion, but requiring that if a galaxy without
redshift is involved in a pair, its absolute magnitude (assigning it
the redshift of the other object) be consistent with the absolute
magnitude limits being used (i.e. it must lie between the selection
lines shown in Figure 1). This yields an extra 36 companions for the
$-22 < M_B < -19$ sample and 47 companions for the $-21 < M_B < -18$
sample. Since some of these pairs may be due to chance superposition,
we repeat our analysis using only galaxies with redshifts, but solely
using the $r_p$ criterion. We find a total of 193 and 262 companions,
respectively, of which we know 137 and 176 to be 'real'. The fraction
of `true' companions is therefore 70\% and we apply this to derive an extra
25.2 companions for the first sample and 32.9 companions for the second
sample. This yields a correction of 18.4\% and 18.7\% to the derived
values of $N_c$ and $L_c$ (for the brighter and fainter sample, respectively) 
owing to galaxies whose redshifts are missing from the survey. 

These values are larger than one might expect from the low redshift 
incompleteness ($\sim 4\%$) of the survey; the main source of incompleteness 
lies in galaxies that were never targeted and these objects lie preferentially 
at close angular separations to other galaxies: since the MGC derives 
most of its redshifts from 2dF observations (including 2dFGRS data) and 
SDSS surveys, the corrections are affected by the instrumental bias against 
close pairs due to fiber placement constraints (although this is lower than 
in the 2dFGRS because of the higher MGC completeness).

\section{Discussion}

After applying the incompleteness correction derived above we find
that $N_c=0.0174 \pm 0.0015$ and $L_c=(252 \pm 30) \times 10^6$
$L_{\odot}$ for galaxies with $-22 < M_B < -19$ and $N_c=0.0357 \pm
0.0027$ and $L_c= (294 \pm 31) \times 10^6 \; L_{\odot}$ for
galaxies with $-21 < M_B < -18$, with mean redshifts as given above. 
If we assume that the merger rate evolves as $(1+z)^m$ with $m \sim 3$ 
\citep{lefevre00,gottlober01,patton02} we find that our $N_c$ of $0.0357 
\pm 0.0027$ translates to $0.0257 \pm 0.0019$ at the mean SSRS2 redshift 
of 0.015, which is in good agreement with the number of companions of 
$0.0225 \pm 0.0052$ measured by P00 for galaxies with $-21 < M_B <
-18$. At $z=0.289$, P02 find $N_c=0.0334 \pm 0.0081$ for these same galaxies, 
converted to our cosmology (D. Patton, private communication), which 
corresponds to $N_c=0.0217 \pm 0.053$  at our mean $z=0.116$. 

\subsection{Comparison with theoretical models}

Predictions of the expected merger rate in CDM models can be derived
from numerical simulations. However, these simulations are usually
conducted at much lower resolution than the typical pair separation
used here. Although it is possible to extrapolate to higher
resolution, by resampling the larger simulations, this procedure is
suspicious as information may already have been lost on small scales
\citep{gottlober01}. Additionally, it is difficult to relate
dark haloes to their luminous content: it is generally assumed that
each galaxy corresponds to a single halo, but this assumption is
likely to be too simplistic.

\citet{khochfar01} calculate the fraction of close pairs and the 
evolution of the merger rate in a $\Lambda$CDM cosmology. The merger 
fraction they derive for galaxies of $M \approx 3 \times 10^{12} M_{\odot}$ 
(which corresponds to $-21 < M_B < -18$) is approximately in agreement 
with our result and P00, but their models cannot fully reproduce 
the drop in the merger fraction by a factor of 2 for the brighter galaxies. 
The simple 'chance hypothesis', where galaxies fall into each other's 
gravitational influence zone simply by random motions, appears to work 
somewhat better: galaxies with $-22 < M_B < -19$ have an approximately 6 
times lower chance to be in close proximity than galaxies with $-21 < M_B 
< -18$, by integration of the luminosity function, but such objects are 
also $\sim 3$ times more clustered \citep{norberg02}. Therefore the pair 
fraction should be lower by about a factor of 2 which is in reasonable 
agreement with our findings. This is apparently in contrast with the 
conclusions of \citet{xu04} who appear to reject the chance hypothesis 
for dwarfs in the proximity of giants; the hypothesis may only apply 
to giant galaxies, which are more resilient than dwarfs. Clearly, a 
more consistent interpretation of these results must await a more 
thorough theoretical modelling of this phenomenon.

Since most companions are found in pairs, rather than in triplets or
higher order $N$-tuples, our derived value of $N_c$ is comparable to
the fraction of galaxies in close pairs. Following \citet{yee95} and P00
we now assume that half of our {\em dynamically} close pairs are actually 
{\em physically} close pairs and will hence merge. We now integrate the 
merger rate to $z=1$ to derive the fraction of present day galaxies that
have undergone a major merger since this time. The fraction of merger 
remnants is (P00):

\begin{equation}
f_{rem}= 1-{\prod_{k=1}^N} {{1-F_{mg} (z_k)} \over {1-0.5 F_{mg} (z_k)}}
\end{equation}

\noindent where $F_{mg} (z)$ is the merger rate at redshift $z$, which is 
assumed to vary as $F_{mg} (z) = F_{mg} (0) (1+z)^m$, with $m=3$, $z_k$ 
corresponds to a lookback time of $t=kT_{mg}$ and $T_{mg}$ is the merger 
timescale. Following P00, we assume that this merger timescale is 0.5 Gyr 
for galaxies with $-21 < M_B < -18$ and, scaling by the luminosity of the galaxies, 
0.2 Gyr for galaxies with $-22 < M_B < -19$. We therefore have 27 merger timescales 
to $z \sim 1$ for the brighter sample and 11 timescales for the fainter sample.  
This implies that $\sim 22.7 \%$ of $-22 < M_B < -19$ galaxies and $ \sim 19.2 \% $
of $-21 < M_B < -18$ galaxies have undergone a major merger since $z \sim 1$, 
approximately the last half of the Hubble time. P00 derive a remnant fraction 
$\sim 6.6\%$ for their sample, but they assume a slightly different cosmology 
and $m=0$, while \citet{lin04} also obtain a remnant fraction of $\sim 10$\%, but use 
$m=0.51$ and P02 derived an integrated merger fraction of 15\% with $m=2.3$. Note,
however, that these values are strongly dependent on the assumed merger timescales, 
index for the evolution of the merger rate and fraction of merging galaxies. 
For the sake of comparison, $\Lambda$CDM models predict that about 50\% of 
$L > L^*$ galaxies have undergone a major merger since $z \sim 1$
\citep{murali02}, which is about a factor of 2 lower than our 
estimate.

\subsection{The properties of galaxies in pairs}

It is interesting to compare the properties of galaxies in pairs with
respect to their parent sample. Since only $ \sim 20\%$ of these 
objects have undergone a merger in the last half of the Hubble time,
this should provide a sample of `undisturbed' galaxies and offer some 
insight as to the effects of close interactions (and mergers) on galaxy 
morphology and star formation. Figure 2 already offers a hint of this: 
many galaxies show obvious signs of interactions, with tidal tails, rings
and other signs of morphological disturbance, a connection exploited
by the asymmetry parameter \citep{conselice03}. 

Figure 3 shows the colour distribution in rest $u-r$ for galaxies in both
samples, compared with that of their parent distribution as well as
the distribution of morphologies in the pairs and parent samples.
Although the errors on the distributions for paired galaxies are
large, because of small number statistics, there is some evidence that
galaxies in pairs tend to be bluer than those in the parent sample,
by approximately 0.1 mag in $u-r$, suggesting that close interactions 
induce star formation episodes \citep{alonso04,nikolic04,barton03,lambas03,
barton00}, although the average blueing for our sample is more consistent
with relatively mild starbursts \citep{bergvall03}. On the other hand, 
not all of our kinematic pairs may be real merging systems and
therefore, not all objects may show enhanced star formation. 

Comparison of the morphologies (determined from visual examination of the
images) shows that there are marginally more E/S0 galaxies and Sd/Irr galaxies 
among pairs, while there are fewer Sabc galaxies. The excess of early-type
galaxies in pairs was noted by \cite{junqueira98}, and this was attributed to
an excess of lenticulars by \citet{rampazzo92} and \citet{rampazzo95} who
suggest that pairs containing early-type members may be the result of early 
merging of groups (which would have additional companions handy to fuel 
further merging). The excess of very late-type galaxies may be explained by 
interactions increasing the star formation rate and therefore leading to later 
classifications. These may be long-lived interacting systems rather than objects 
in an early phase of a merger \citep{junqueira98}.  

Mergers and interactions therefore alter the morphologies of the member galaxies
and induce star formation episodes. Scenarios in which local conditions (i.e.
mergers and close encounters) are primarily responsible for morphological
evolution and the decline of star formation in denser environments have been
recently proposed following analysis of large local samples of field and
cluster galaxies \citep{balogh04,depropris04,croton05}, although the change
in the fraction of blue galaxies is much higher than the merger fraction we
derive (suggesting a large contribution from lower luminosity galaxies). Wider
and deeper surveys would be valuable, in allowing consideration of the
importance of minor mergers.

\acknowledgments

RDP is supported by a PPARC fellowship. JL acknowledges an ESO
fellowship. The Millennium Galaxy Catalogue consists of imaging data 
from the Isaac Newton Telescope and spectroscopic data from the Anglo 
Australian Telescope, the ANU 2.3m, the ESO New Technology Telescope, 
the Telescopio Nazionale Galileo, and the Gemini Telescope. The survey 
has been supported through grants from the Particle Physics and Astronomy 
Research Council (UK) and the Australian Research Council (AUS). The data 
and data products are publicly available from http://www.eso.org/$\sim$jliske/mgc/ 
or on request from JL or SPD. We would like to thank the referee, D. R.
Patton, for a number of helpful comments and suggestions that made the
paper clearer and better.

\clearpage


\begin{figure}
\epsscale{0.9}
\plotone{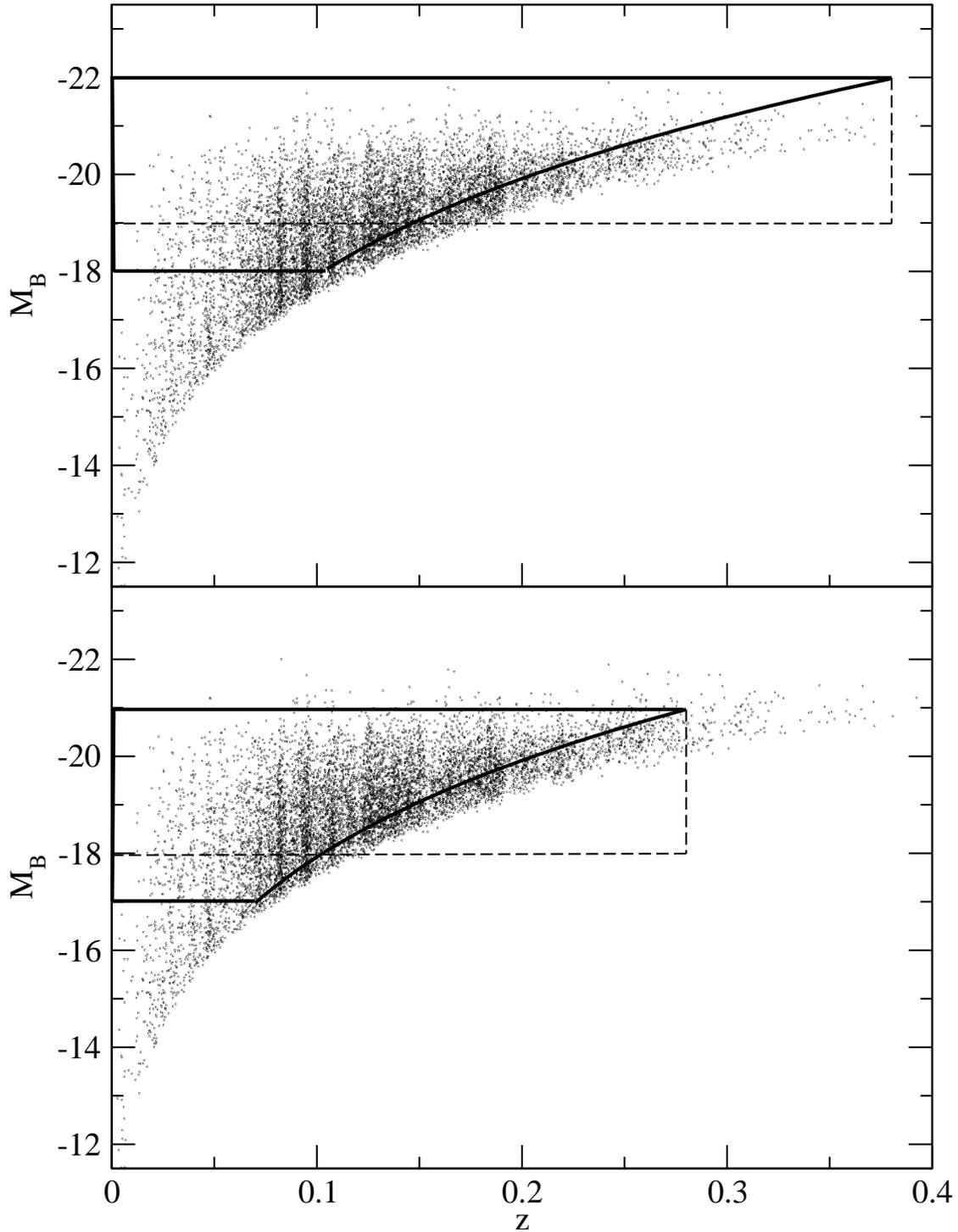}
\caption{The absolute magnitude-redshift distribution of MGC
galaxies, showing the selection lines for the $-22 < M_B < -19$
and the $-21 < M_B < -18$ samples in the upper and lower panel,
respectively. In each panel, the thick solid line 
delimits the magnitude limited sample with $M_{bright} < M_B < M_{lim}(z)$
used in the analysis, while the dashed box shows the hypothetical 
volume limited sample with $M_{bright} < M_B < M_2$ to which the 
results are normalized.}
\end{figure}

\clearpage 

\begin{figure}
\plotone{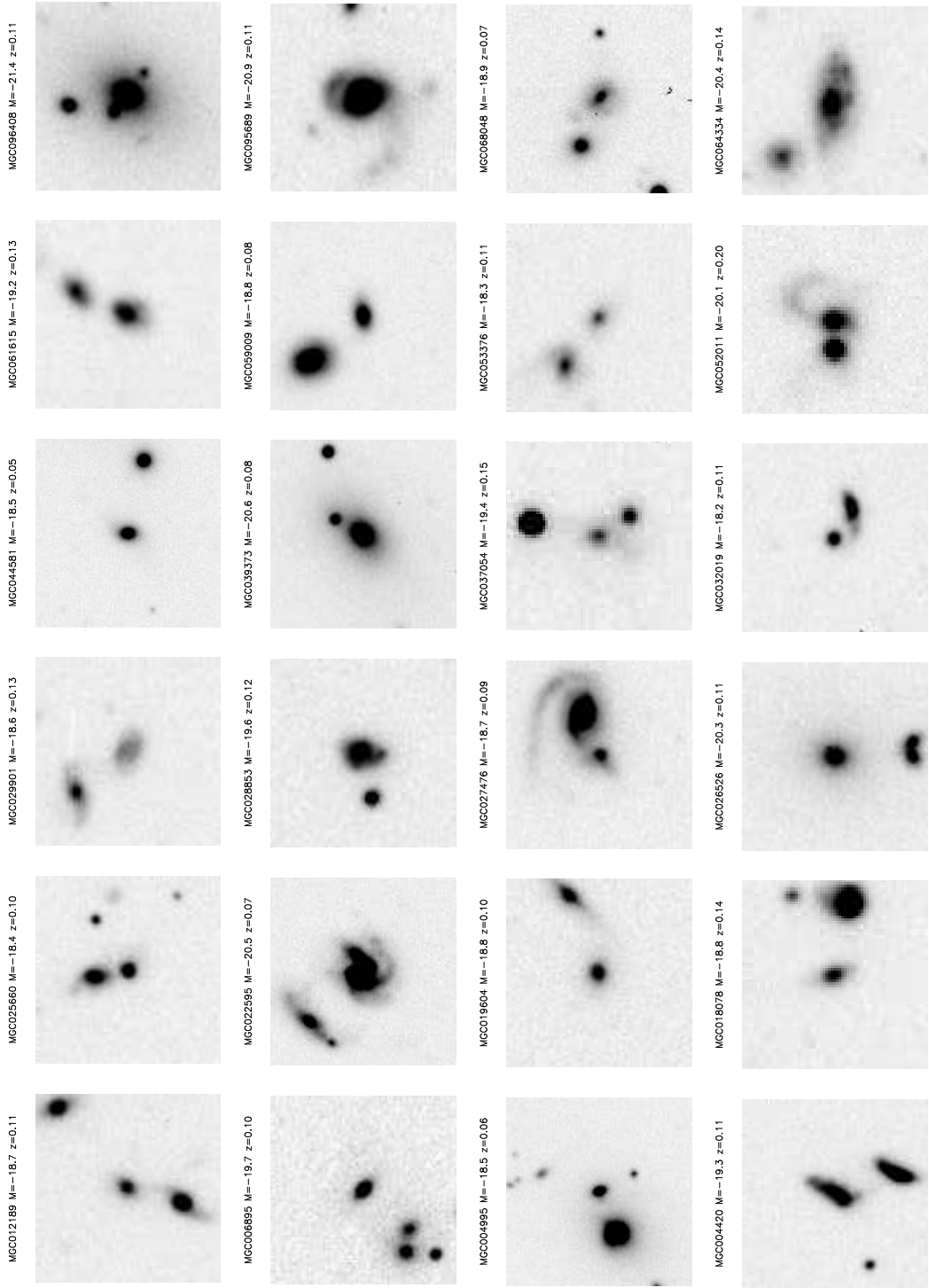}
\caption{Postage stamps of a random selection of pair galaxies. The stamps
are $40 h^{-1}$~kpc  on the side and are centered on a randomly chosen pair 
member.}
\end{figure}

\clearpage

\begin{figure}
\plotone{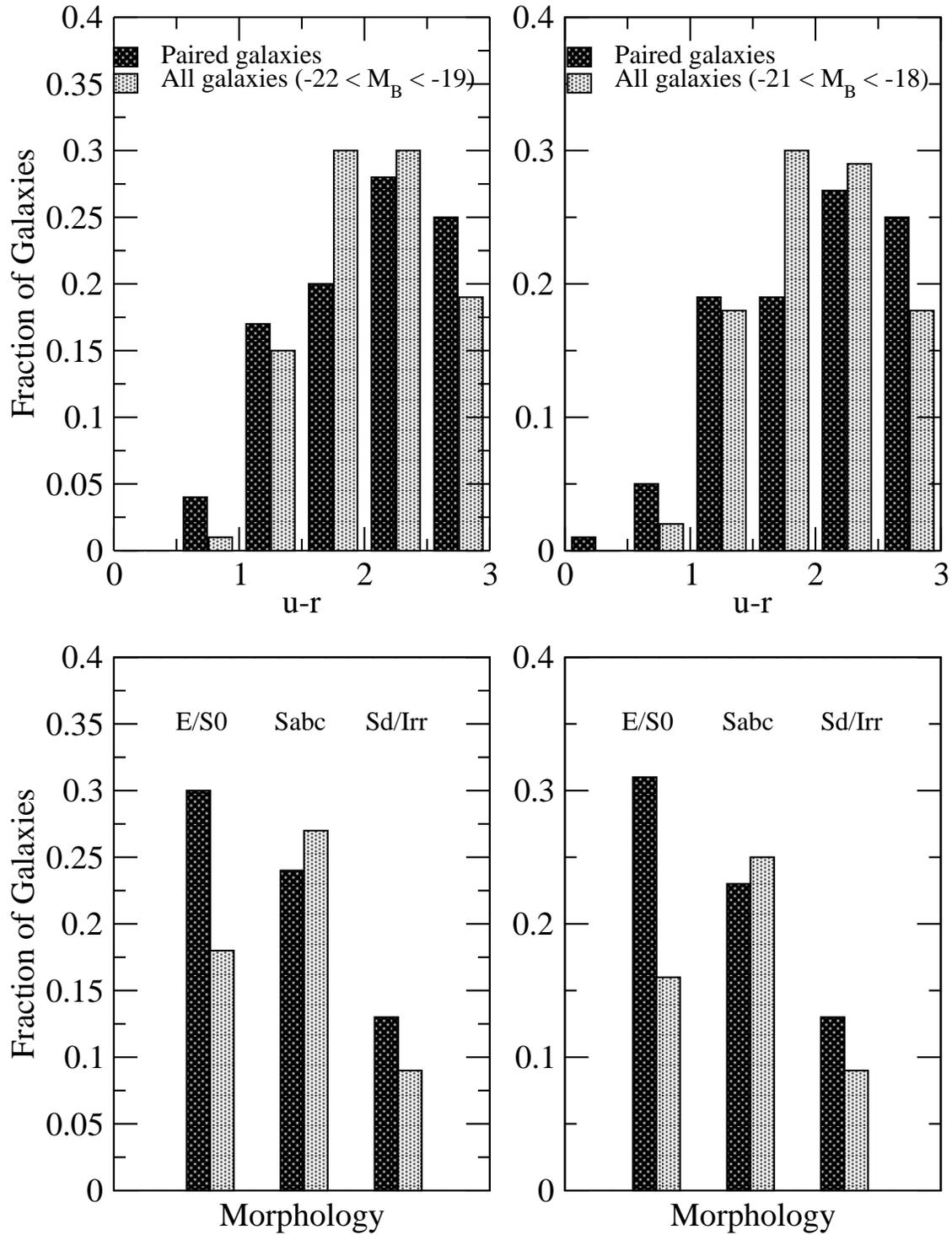}
\caption{Histograms of the colour and morphology distributions for galaxies in
pairs and for their parent samples.}
\end{figure}

\clearpage



\clearpage

\begin{deluxetable}{lcccc}
\tablecaption{MGC: Unique close pairs and triples}
\tabletypesize{\small}
\tablewidth{0pt}
\tablehead{
\colhead{MGC ID} & \colhead{Comp. ID} &\colhead{RA (2000)} & \colhead {Dec. (2000)} & \colhead{$cz$} 
}
\startdata
533 & 537 & 10:00:23.9 &  00:00:51.9 & 18918 \\
919 & 921 & 10:01:17.9 & -00:11:02.9 & 27348 \\
956 & 960,963 & 10:01:28.3 & -00:12:44.1 & 18525 \\
973 & 978 & 10:01:08.3 & -00:13:32.7 & 28017 \\
977 & 95002 & 10:01:02.9 & -00:13:38.8 & 27411 \\
1579 & 1580 & 10:03:44.2 & 00:01:50.5 & 19611 \\
2858 & 2866 & 10:08:56.4 & -00:02:16.5 & 28737 \\
3118 & 3121 & 10:09:03.8 & -00:16:42.8 & 20574 \\
3526 & 3530 & 10:11:21.4 & -00:10:07.3 & 20907 \\
4397 & 4742 & 10:15:53.3 & 00:05:07.4 & 22011 \\
4420 & 4421 & 10:16:28.6 & -00:01:34.1 & 31554 \\
4450 & 4452 & 10:15:44.1 & -00:03:02.7 & 42861 \\
4484 & 4486 & 10:16:23.3 & -00:04:49.7 & 28458 \\
4651 & 4655 & 10:16:25.1 & -00:14:25.2 & 21162 \\
4988 & 4995 & 10:17:48.6 & -00:01:50.3 & 18621 \\
5974 & 5977 & 10:22:07.5 & -00:03:39.6 & 21354 \\
6551 & 6552 & 10:22:58.6 & -00:10:23.7 & 33912 \\
6610 & 6614 & 10:23:35.5 & -00:13:17.3 & 28611 \\
6888 & 6895 & 10:26:13.8 & -00:02:00.8 & 31461 \\
6892 & 6895 & 10:26:13.9 & -00:02:04.3 & 31029 \\
7876 & 7881 & 10:28:56.3 & -00:02:02.0 & 28410 \\
8169 & 8170 & 10:30:11.0 & 00:00:51.2 & 33249 \\
12189 & 12190 & 10:49:20.1 & 00:14:17.8 & 33327 \\
13718 & 13720 & 10:54:50.9 & -00:09:16.4 & 32703 \\
15486 & 15490 & 11:03:31.6 & -00:10:27.1 & 22404 \\
16242 & 16245 & 11:07:38.2 & -00:01:59.2 & 19728 \\
16245 & 16250 & 11:07:38.2 & -00:02:13.1 & 19776 \\
16408 & 16413 & 11:07:44.9 & -00:11:21.5 & 20094 \\
17921 & 18001 & 11:15:22.0 & 00:06:55.2 & 8193 \\
18822 & 18824 & 11:17:50.8 & 00:03:07.6 & 29955 \\
19025 & 19318 & 11:19:51.1 & 00:05:12.4 & 31302 \\
19565 & 19579 & 11:20:57.1 & -00:05:03.7 & 7506 \\
19604 & 19605 & 11:20:53.2 & -00:07:29.6 & 29520 \\
21115 & 21118 & 11:29:46.1 & 00:14:33.1 & 31263 \\
22241 & 22246 & 11:32:40.1 & 00:01:35.7 & 39351 \\
22591 & 22594 & 11:35:55.1 & -00:16:07.4 & 19590 \\
22594 & 22595 & 11:35:56.6 & -00:16:12:0 & 19698 \\
23246 & 23247 & 11:40:15.9 & -00:09:59.2 & 22626 \\
25070 & 25072 & 11:49:47.9 & 00:07:42.7 & 23514 \\
25500 & 25502 & 11:51:36.3 & 00:00:02.2 & 18150 \\
25601 & 25604 & 11:52:26.6 & -00:06:30.2 & 38721 \\
25660 & 25662 & 11:50:51.4 & -00:10:53.7 & 31017 \\
25897 & 25900 & 11:54:31.3 & 00:10:20.2 & 32496 \\
26507 & 96408 & 11:56:19.5 & -00:12:18.3 & 32637 \\
26526 & 26527 & 11:56:01.7 & -00:13:04.7 & 32535 \\
26724 & 26725 & 11:57:37.6 & 00:07:32.0 & 32235 \\
26970 & 26973 & 11:56:37.8 & -00:11:10.0 & 31950 \\
27383 & 27386 & 12:00:28.9 & -00:07:24.5 & 24288 \\
27476 & 27477 & 11:58:41.6 & -00:12:47.6 & 28482 \\
27627 & 27629 & 12:01:46.0 & 00:14:32.2 & 31470 \\
27897 & 27898 & 12:00:44.7 & -00:09:22.2 & 49578 \\
29266 & 29269 & 12:06:43.0 & -00:12:13.7 & 27825 \\
29405 & 29407 & 12:09:41.2 & 00:12:06.2 & 29826 \\
29901 & 29904 & 12:12:09.9 & 00:10:29.7 & 37899 \\
31392 & 31394 & 12:16:58.5 & -00:13:18.4 & 21576 \\
31991 & 95166 & 12:21:56.5 & 00:09:48.7 & 31875 \\
32019 & 32022 & 12:21:48.2 & -00:00:23.4 & 32475 \\
32110 & 32111 & 12:22:17.9 & -00:07:43.1 & 51849 \\
33182 & 33189 & 12:27:22.7 & 00:06:43.2 & 47955 \\
35114 & 35124 & 12:35:42.4 & -00:12:53.9 & 6873 \\
37672 & 37676 & 12:47:42.5 & -00:08:14.3 & 26871 \\
37916 & 37918 & 12:50:28.1 & 00:13:36.3 & 13989 \\
37988 & 37990 & 12:50:31.9 & 00:09:30.3 & 24855 \\
39047 & 39048 & 12:53:02.0 & -00:16:12.6 & 25461 \\
39373 & 39375 & 12:54:37.8 & -00:11:04.4 & 24822 \\
40301 & 40302 & 13:02:21.9 & 00:14:40.3 & 20403 \\
41961 & 41965 & 13:07:02.4 & 00:03:24.2 & 23976 \\
44577 & 44581 & 13:18:41.2 & -00:10:25.8 & 14487 \\
45563 & 45565 & 13:22:47.1 & -00:13:47.4 & 24699 \\
45798 & 45804 & 13:26:34.4 & 00:09:09.8 & 25878 \\
46553 & 45565 & 13:27:49.0 & -00:15:34.1 & 42756 \\
49445 & 49450 & 13:40:17.2 & 00:02:06.0 & 43479 \\
52736 & 52742 & 13:53:26.2 & -00:09:04.2 & 31308 \\
52767 & 52768 & 13:53:18.7 & -00:10:16.1 & 56688 \\
53371 & 53376 & 13:55:41.2 & -00:14:44.5 & 31623 \\
53554 & 53560 & 13:58:41.9 & 00:14:51.5 & 9942 \\
53797 & 95689 & 13:56:46.2 & -00:08:54.1 & 32229 \\
56147 & 56149 & 14:05:02.9 & -00:15:05.0 & 9777 \\
59005 & 59009 & 14:16:17.2 & -00:09:25.2 & 25059 \\
59318 & 59324 & 14:17:59.6 & 00:16:02.8 & 15963 \\
59445 & 59446 & 14:16:45.1 & -00:01:14.3 & 37581 \\
59700 & 59721 & 14:17:37.2 & 00:03:50.6 & 15882 \\
60716 & 60719 & 14:20:56.7 & -00:04:23.6 & 30654 \\
60869 & 60904 & 14:22:27.5 & 00:03:38.5 & 9729 \\
61615 & 61618 & 14:24:10.7 & 00:01:47.2 & 37986 \\
64206 & 64334 & 14:34:21.9 & 00:06:27.2 & 41031 \\
64512 & 64513 & 14:33:10.6 & -00:08:07.2 & 40839 \\
64512 & 64515 & 14:33:10.6 & -00:08:07.2 & 40839 \\
68038 & 68048 & 14:44:04.5 & 00:03:14.6 & 22059 \\
95060 & 95061 & 12:08:11.7 & 00:03:50.2 & 29916 \\
96246 & 96247 & 14:27:30.7 & -00:10:38.6 & 48996 \\
96944 & 96945 & 14:15:24.6 & 00:16:02.2 & 53166 \\
\enddata
\end{deluxetable}

\end{document}